\RequirePackage{lineno}
\documentclass[aps, prl,
floatfix,twocolumn,superscriptaddress,
]{revtex4-1}
\pdfoutput=1
\usepackage{t1enc}
\usepackage[utf8]{inputenc}
\usepackage[T1]{fontenc}
\usepackage{lmodern}
\usepackage{graphicx}
\usepackage{epsfig}
\usepackage{amssymb}
\usepackage{amsmath}
\usepackage{dcolumn}
\usepackage{bm}
\usepackage{color}
\usepackage{float}
\usepackage{hyperref}
\usepackage{natbib}
\usepackage{footnote}
\usepackage{ulem}
\usepackage{hhline}
\begin{document}

\title{Pb-apatite framework as a generator of novel flat-band CuO based physics}

\author{Rafa{\l}\,Kurleto}
\email{Rafal.Kurleto@Colorado.edu}
\affiliation{Department of Physics, University of Colorado, Boulder, CO, 80309, USA}
\affiliation{Center for Experiments in Quantum Materials, University of Colorado Boulder, Boulder, CO 80309}

\author{Stephan\,Lany}
\affiliation{National Renewable Energy Laboratory, Golden, Colorado 80401}

\author{Dimitar\,Pashov}
\affiliation{King's College London, Theory and Simulation of Condensed Matter, The Strand, WC2R 2LS London, UK}

\author{Swagata\,Acharya}
\affiliation{National Renewable Energy Laboratory, Golden, Colorado 80401}

\author{Mark\,van Schilfgaarde}
\email{Mark.vanschilfgaarde@gmail.com}
\affiliation{Department of Physics, University of Colorado, Boulder, CO, 80309, USA}
\affiliation{National Renewable Energy Laboratory, Golden, Colorado 80401}

\author{Daniel S.\,Dessau}
\email{Dessau@Colorado.edu}
\affiliation{Department of Physics, University of Colorado, Boulder, CO, 80309, USA}
\affiliation{Center for Experiments in Quantum Materials, University of Colorado Boulder, Boulder, CO 80309}
\affiliation{National Renewable Energy Laboratory, Golden, Colorado 80401}

\begin{abstract}\
Based upon density functional theory (DFT) calculations, we present the basic electronic structure of
CuPb$_9$(PO$_4$)$_6$O (Cu-doped lead apatite, aka LK-99), in two scenarios: (1) where the structure is constrained to the P3 symmetry and (2) where no symmetry is imposed. At the DFT level, the former is predicted to be metallic while the latter is found to be a charge-transfer insulator. In both cases the filling of these states is nominally~$d^9$, consistent with the standard Cu$^{2+}$ valence state, and Cu with a local magnetic moment of order $0.7$~$\mu_B$. In the metallic case we find these states to be unusually flat ($\sim$0.2~eV dispersion), giving a very high density of electronic states (DOS) at the Fermi level that we argue can be a host for novel electronic physics, including potentially high temperature superconductivity. The flatness of the bands is the likely origin of symmetry-lowering gapping possibilities that would remove the spectral weight from $E_F$. Since some experimental observations show metallic or semiconducting (see https://arxiv.org/abs/2308.05001) behavior, we propose that disorder (likely structural)  is responsible for closing the gap.  We consider a variety of possibilities that could possibly close the charge-transfer gap, but limit consideration to kinds of disorder that preserve electron count.  Of the possible kinds we considered (spin disorder, O populating vacancy sites, Cu on less energetically favorable Pb sites), the local Cu moment, and consequently the charge-transfer gap remains robust.  We conclude that disorder responsible for metallic behavior entails some kind of doping where the electron count changes. Further, we claim that the emergence of the flat bands should be due to weak wave function overlap between the orbitals on Cu and O sites, owing to the directional character of the constituent orbitals. Therefore, finding an appropriate host structure for minimizing hybridization between Cu and O while allowing them to still weakly interact should be a promising route for generating flat bands at $E_F$ which can lead to interesting electronic phenomena, regardless of whether LK-99 is a room-temperature superconductor.

\end{abstract}


\maketitle

\section{Introduction}

A brand-new potential superconductor with chemical formula Cu$_x$Pb$_{10-x}$(PO$_4$)$_6$O ($0.9<x<1.1$) named “LK-99”
was recently reported~\cite{lee_2023,lee_2023_1} claiming to be the world’s first ambient pressure room-temperature
superconductor. In particular, many signatures of superconductivity were found to exist up to 400~K (127$^{\circ}$C),
including diamagnetic responses and strong reductions in electrical resistance.  And while skepticism as to the validity
of the claims of superconductivity is strong and will likely exist until the experimental findings are verified by a few
other groups, here we show that the electronic structure of this class of compounds is very interesting in its own
right, and has many features expected to be supportive of superconductivity at high temperatures. In particular, we show that, within DFT, the process of doping Cu atoms on the particular Pb(4\emph{f}) site described in
Refs.~\cite{lee_2023,lee_2023_1} generates either a metal, when the system is constrained to keep
  \emph{P}3 symmetry as reported, or an insulator when the system is allowed to fully relax.
  The orbital character of states around the Fermi level $E_{F}$ also depends on whether relaxation is constrained or
  not, as we explain below.  In either case two bands near $E_{F}$ appear, consisting predominately of Cu \emph{d} and O
  \emph{p} character and have an especially small bandwidth of order 0.2~eV.  In the symmetry-constrained case, the
  three states coalesce and the system is metallic in DFT and DFT+U (U=5 eV), while in the fully relaxed case, a state of Cu \emph{d} state with $|m|{=}1$ in one spin channel splits off and becomes unoccupied. Thus Cu becomes $d^{9}$ and the system becomes
  insulating.  In both cases the local Cu moment is ${\sim}0.7\mu_B$.

Experience with many other electronic compounds with flat electronic bands right at the Fermi level immediately signals
the possibility of exotic properties as such states tend to be extremely “active”.  That is, the high density of
electronic states at $E_F$ that comes with these flat bands tends to lead to electronic instabilities of various kinds, including charge density waves and/or superconductivity. Also helpful for generating exotic properties are systems with
small spin magnetic moments as these usually bring strong quantum fluctuations of the spins. A good example is the Cu-O
based “cuprate” high temperature superconductors~\cite{bednorz_1986} with spin 1/2 moments on the Cu sites, or more
specifically, partially delocalized across Cu-O clusters~\cite{zhang_1988}.  We find that the present system also should
contain spin 1/2 moments associated with Cu, consistent with the $d^9$ (Cu$^{2+}$) valence state that our calculations
show.  Indeed, this system has many similarities with traditional cuprate superconductors, as we explain below.

\begin{figure}
\centering\includegraphics[width=0.9\linewidth]{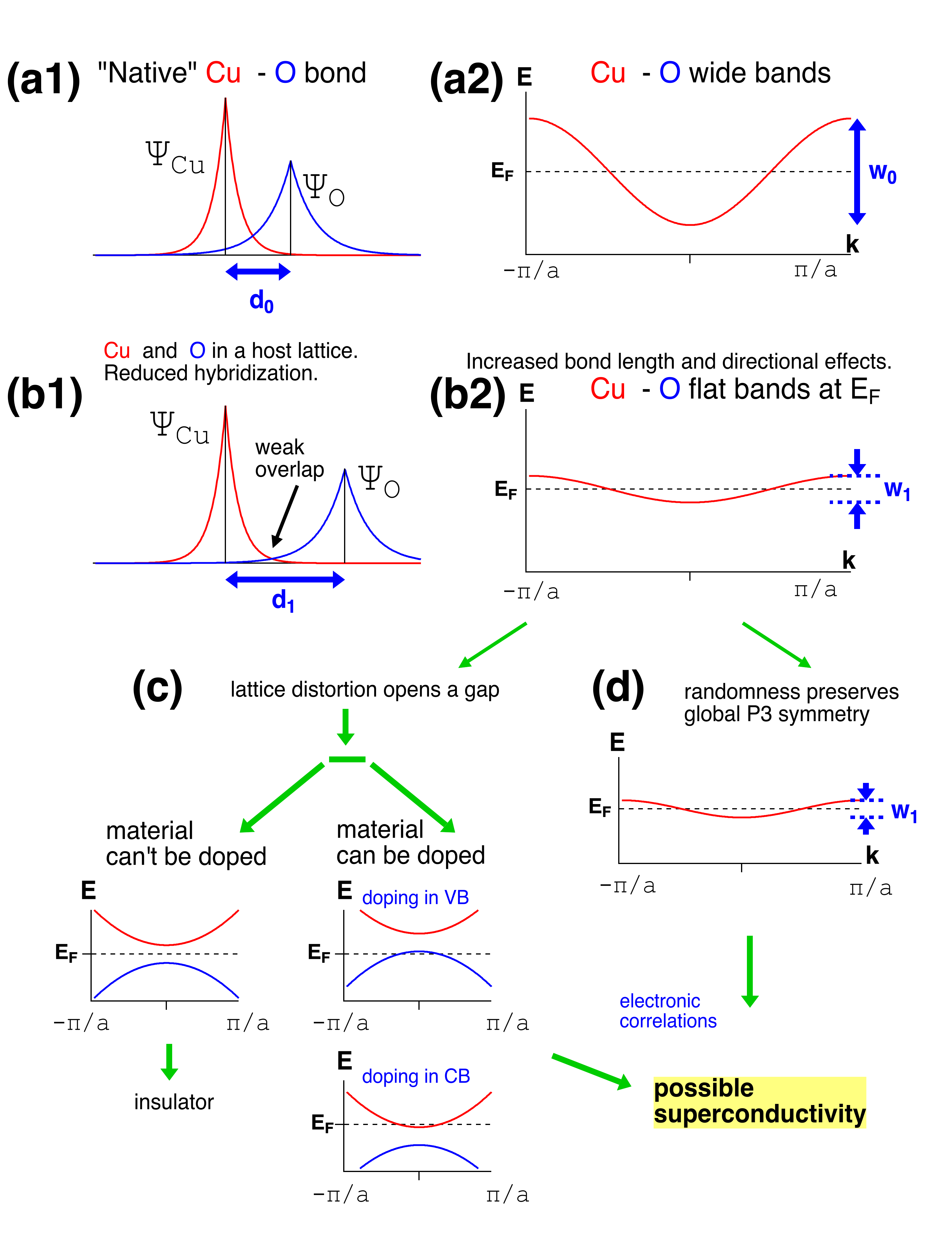}
\caption{Roadmap for obtaining strongly renormalized ultra-flat Cu-O bands in a host lattice, potentially leading to high temperature superconductivity. Our calculations show that we are in between the steps of panels (b) and (c).}
\label{fig:1}
\end{figure}

We will argue that these flat Cu-O derived bands do not occur by accident in Cu$_x$Pb$_{10-x}$(PO$_4$)$_6$O ($0.9<x<1.1$), but rather should be due to minimal direct coupling between the Cu 3d and O 2p wave functions.  (Already in the parent compound, the band dispersion of the O-derived valence band maximum itself has narrow dispersion.) This reduced coupling gives a reduced $E$ vs. $k$ band dispersion (flat bands) and an associated high electronic density of states. We find a portion of these flat bands/ high density of states live right at $E_F$ so they should become strongly electronically “active” – potentially generating superconductivity at relatively high temperatures through one of many possible channels, as will be discussed later. If superconductivity does indeed exist in LK-99, our results imply that it is enabled by the ultra-flat Cu-O bands at $E_F$.  If it does not exist in LK-99, we suggest that the concepts outlined here can be used as guidance for finding other possible hosts besides Pb apatite that generate electronically active flat bands at $E_F$. Fig.~\ref{fig:1} shows a schematic of how to obtain ultra-flat bands, as illustrated for Cu~(red) and O~(blue) hybridized wave functions. Panel (a1) is illustrative of "native" Cu-O bonds, where the wave function overlap is large. In this case (as occurs in the cuprates) Cu and O form strong covalent bonds, which gives rise to a significant dispersive bandwidth $w_0$ (a2).  (Panel b1)  The Cu and O atoms only weakly hybridize when Cu is placed in the Pb-apatite framework giving a small bandwidth from these states. Note that the phosphate (PO$_4$ unit) indeed forms strong covalent bonds but these bonding states reside far from the Fermi level. 

 In a metal with very flat bands bands at $E_F$, the system would typically like to find some way to lower the symmetry and open gaps (panel c). The flatter the bands get, the greater the driving force to form a gap.  There are many possibilities, including  structural relaxation that breaks the P3 symmetry (see below); with this symmetry-lowering gapping strongly competing with superconductivity, as the superconductivity requires the system to be metallic.  Panel (d) illustrates how small amounts of random disorder, for example from incommensurate doping levels ($x{\neq}1$) or doping onto different nominally equivalent sites, can protect the flat bands from such gapping effects, keeping them available to support high temperature superconductivity.  We note that this system, owing to its weak hybridization, may be more sensitive to disorder than established superconductors.  In this regard we note that non-sign changing s-wave superconductivity should generally be robust against the same disorder that can be used to protect the flat bands from other gapping mechanisms, as first shown by Anderson~\cite{anderson_1959}.

These ingredients are different from those proposed in Refs~\cite{lee_2023,lee_2023_1} for the superconductivity.  They
argue that the relevant states for the superconductivity are $6s^1$ states from the six-fold degenerate Pb atoms (which
we refer to as Pb(6\emph{h}) following Wyckoff notation for the original Pb-apatite crystal structure)
with these states affected by the compressive strain of the inserted Cu atoms, leading to an insulator to metal
transition and setting the stage for superconductivity.  Our electronic structure calculations show negligible amounts
of Pb 6$s$ spectral weight near the Fermi level in the symmetry-constrained case (some Pb 6$s$ does appear in the
fully relaxed case).  Contributions originate mostly from the Cu 3\emph{d}, O 2\emph{p} orbitals. Therefore, if there
is superconductivity in this system, it should be from the Cu and O states that are held in place by the apatite
lattice, rather than by the size effect from substituting Cu for Pb.

Following this introduction, we present the details of our electronic structure calculations, beginning with the structures that these calculations are based upon.

\section{Crystal and electronic structure of the parent compound Pb$_{10}$(PO$_4$)$_6$O}

The “parent” compound of the material under consideration is the Pb apatite Pb$_{10}$(PO$_4$)$_6$O. This compound had been previously synthesized in single crystalline form 20 years ago, with detailed structural characterizations from X-ray diffraction measurements performed on these crystals~\cite{krivovichev_2003}, including a CIF structure file that is available in the ICSD database. This structure belongs to the P$_6$3/m hexagonal symmetry class (number 176) and consists of four Pb(4\emph{f}) and six Pb(6\emph{h}) sites, six PO$_4$ tetrahedra each containing one O(1), two O(2), and one O(3) site, as well as four O(4) sites that are each 1/4 occupied. (Often Pb(6\emph{h}) and Pb(4\emph{f}) are denoted as Pb(1) and Pb(2), but the literature is not consistent.) We also note that the 1/4 occupation of the four O(4) sites implies that the experimental characterization describes only the average crystal structure, which does not capture the differences in the local geometries between occupied and unoccupied O(4) sites. To deal with the 1/4 occupation in our electronic structure calculations, we made the approximation that one of the four O(4) sites within the 41 atom primitive cell, hosting one formula unit (fu), should be occupied while the other three are not. This choice lifts the symmetry-equivalence of the O(4) sites and creates a long-range ordered structure with P3 (\#143) space group symmetry. This structure is an imperfect but useful model to study Cu substitution in Pb$_{10}$(PO$_4$)$_6$O, as it is small enough to allow enumeration of all atomic configurations resulting from atomic site substitutions.  

\begin{figure}
\centering\includegraphics[width=1.0\linewidth]{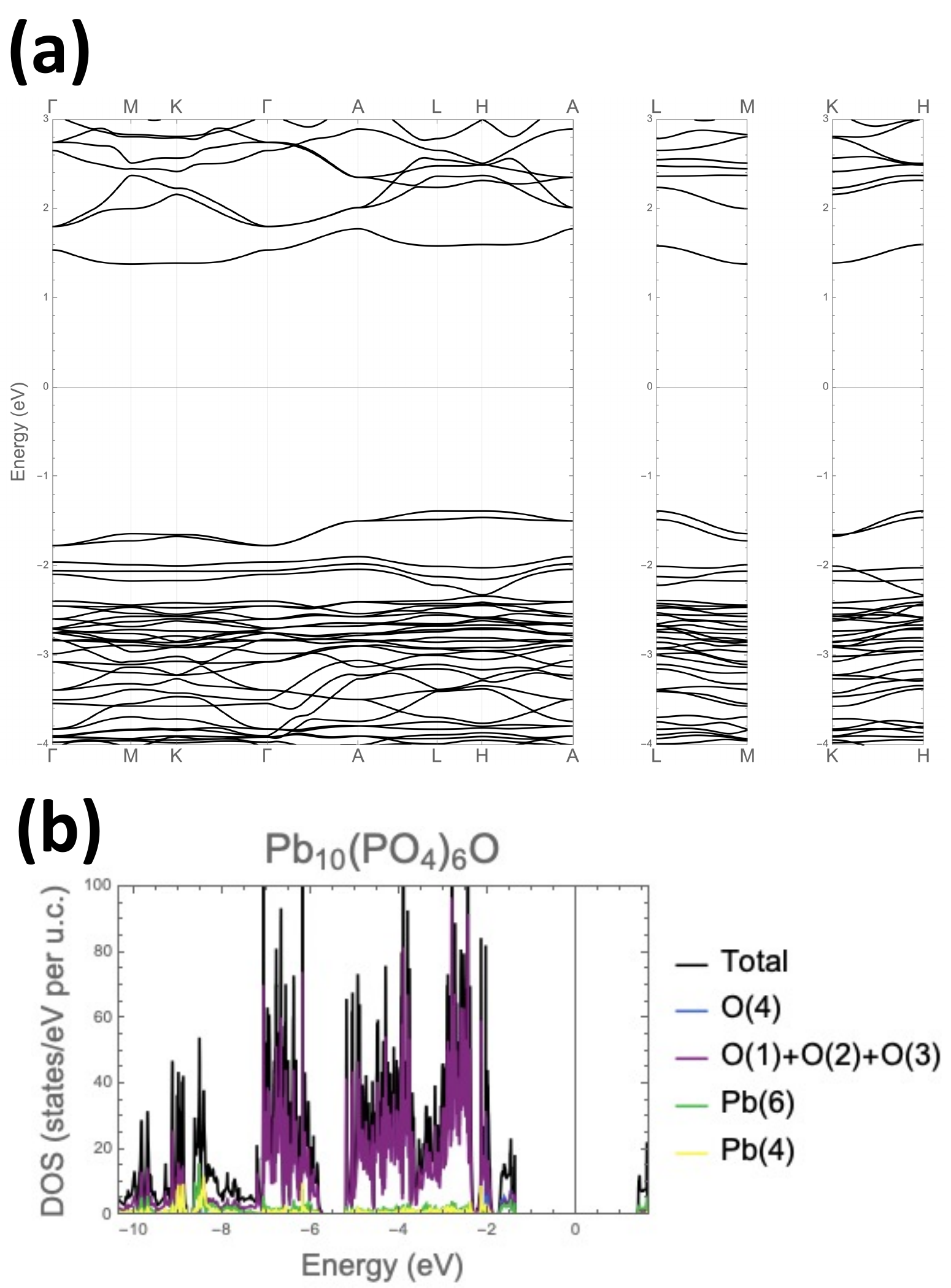}
\caption{Electronic structure of parent compound Pb$_{10}$(PO$_4$)$_6$O. (a) Band dispersion calculated along high symmetry paths. (b) Density of states projected onto different atomic sites. We refer the energies to the Fermi level (consistently within the entire manuscript), which is placed in the middle of the gap. The gap width is equal to 2.765~eV.}
\label{fig:2}
\end{figure}

We calculated the band structure of this compound (see Fig.~\ref{fig:2}) using PBE functional and performed a structural relaxation with the P3 symmetry constrained. We selected one particular position (from four available), consistently with the 4f-1 configuration, which is described later in the text. Panel (a) shows the band dispersion along high symmetry directions in the first Brillouin zone. The indirect energy gap is visible between the top of the valence band at the L point and the minimum of the conduction band at the M point. The width of the gap is equal to 2.765~eV. Calculated partial densities of states allow us to assign main atomic contributions to bands at the edges of the band gap, with the upper edge of the valence band states composed principally of states related to Pb(6h) atoms and oxygen from PO$_4$ tetrahedra, i.e. O(1), O(2), and O(3) together. The distribution of contributions to the DOS at the bottom of the conduction band is more uniform: the dominant contribution comes from the lone oxygen O(4), which is enclosed in Pb(6h) chains. Pb(6h) atoms and the rest of oxygen contribute almost the same number of states as O(4) in this energy region. Overall, the DOS between -8 and -2~eV is dominated by oxygen contribution. Pb(4) is visible as a spike at -2.16~eV which is in coincidence with some flat bands in this region (see panel (b)) and this contribution gets more important between -10 and -8~eV.

\section{Crystal structure of CuPb$_9$(PO$_4$)$_6$O}
We note from the outset that the physical structure of CuPb$_9$(PO$_4$)$_6$O (or the LK-99 materials) has not yet been fully determined. As an alloy system, this material is inherently atomically disordered, and to make the most accurate electronic structure calculations, one needs to determine realistic representative atomic structure models. The present work takes only the first step into this direction, but the variety of atomic and electronic configurations considered here allows us to draw first conclusions about CuPb$_9$(PO$_4$)$_6$O with confidence.

In addition to the 41 atom primitive cell, we also consider a limited number of plausible atomic configurations in supercells containing 82 atoms ($1 \times 1 \times 2$ cell) and 328 atoms ($2 \times 2 \times 2$ cell). Crucially, these cells allow us to study the magnetic interactions between Cu$^{2+}$ ions and to determine the impact of the simplest kinds of disorder on the opening of an energy gap. Fully accounting for atomic and magnetic disorder in an explicit atomistic model (as opposed to an average structure as observed in diffraction experiments) would require configuration sampling in large supercells. Monte-Carlo (MC) simulations of disorder have long been the domain of model Hamiltonian approaches like cluster expansion~\cite{ravi_2012, cordell_2021}, especially for combined sampling of atomic and magnetic configurations, \cite{lavrentiev_2010, decolvenaere_2019}  but DFT-based first-principles MC simulation are becoming increasingly feasible and effective~\cite{gray_2021, rom_2023}. Until representative atomic structure models are obtained from such sampling studies, we believe that the present results can help give a first-order understanding of the novel electronic states that should exist in this compound.

\begin{figure*}
2\centering\includegraphics[width=0.9\linewidth]{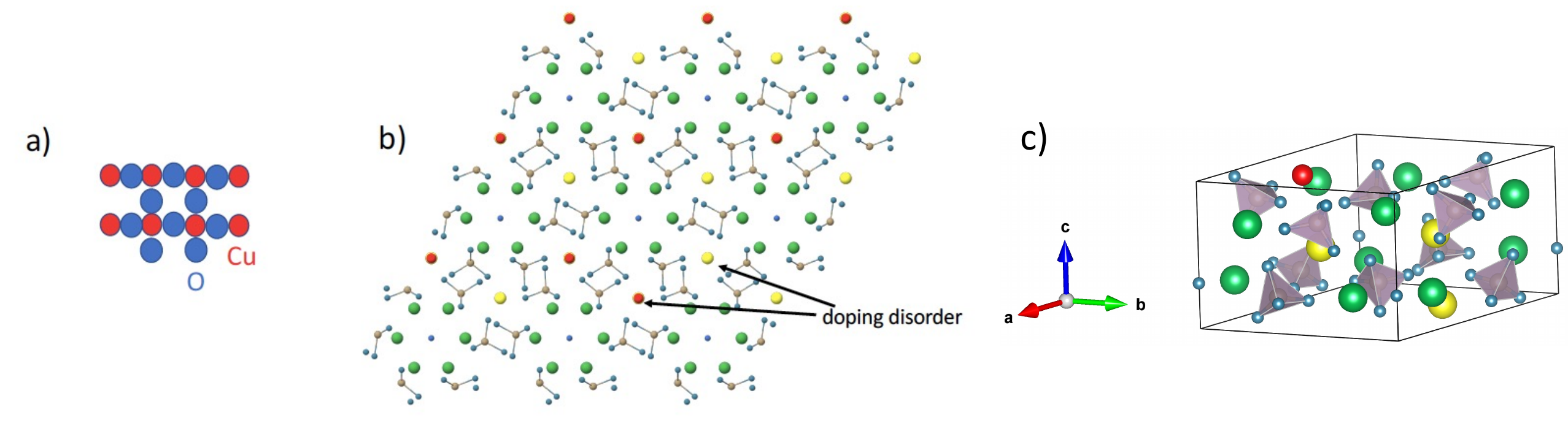}
\caption{(a) Natural bond distances between Cu and O in a Cu-O sublattice, give a large bandwidth $w_0$. (b) Top view projection of CuPb$_9$(PO$_4$)$_6$O focusing on the Cu sites (red) which are substituted on 1/4 (~1/2 in this view) of the four-fold Pb(4\emph{f}) sites (yellow), the blue O(4) atoms that will hybridize with the red coppers, and the six fold Pb(6\emph{h}) (green). Also shown at the bottom right is one example of doping disorder, where the Cu goes into the "wrong" Pb(4f) site.  (c) Shows a 41 atom unit cell in 3D with Cu at selected Pb(4f) position after structural relaxation with no symmetry constraints. PO$_4$ tetrahedra are shaded gray.}
\label{fig:3}
\end{figure*}

We need to note that full ionic relaxation will further reduce the symmetry introducing a weak distortion, with a resulting triclinic (space group P1) crystal structure. However the deviation from the hexagonal unit cell is quite small (the angles between lattice vectors are: $\alpha=89.0003^{\circ}, \beta=89.7000^{\circ},\gamma=121.3085^{\circ}$), and the main structural features are preserved~(Fig.~\ref{fig:3}~c).

References~\cite{lee_2023,lee_2023_1} indicate the composition of the new superconductor as Cu$_x$Pb$_{10-x}$(PO$_4$)$_6$O with ($0.9<x<1.1$) so we choose $x=1$, i.e. CuPb$_9$(PO$_4$)$_6$O, which is commensurate with the 41 atom cell. Reference~\cite{lee_2023_1} states that the Cu replaces one of the four original Pb(4\emph{f}) atoms, which here we confirm through total energy calculations (see below). The Cu substituted structure is illustrated in~Fig.~\ref{fig:3}~b, which shows a top view projection of CuPb$_9$(PO$_4$)$_6$O focusing on the Cu sites (red) which are substituted on 1/4 ($\sim$1/2 in this view) of the four-fold Pb(4\emph{f}) sites (yellow), the blue O(4) atoms, and the six-fold Pb(6\emph{h}) (green). The PO$_4$ tetrahedra are indicated by lines connecting P and O atoms. These structures form strong two-center bonds as noted above, and are electronically inactive at the Fermi level. Also shown at the bottom right is one example of doping disorder, where the Cu goes into the alternative (“wrong”) Pb(4\emph{f}) position. Fig.~\ref{fig:3}~c shows a three-dimensional view of the CuPb$_9$(PO$_4$)$_6$O structure. 

Reference~\cite{lee_2023_1} also indicates that compared to the parent compound the in-plane lattice constant $a$  shrinks from 9.865~{\AA} to 9.843~{\AA} while the out-of-plane lattice constant $c$ shrinks from 7.431~{\AA} to 7.428~{\AA}, i.e. an overall volume shrinkage of 0.48 percent. We determined the lattice vectors by minimizing the total energy using different functionals and obtained good agreement with the experimental lattice constants (see below).

\section{Results: Total energy calculations}

Total energy and atomic relaxations were calculated within density functional theory (DFT) at the levels of the
generalized gradient approximation (GGA)~\cite{perdew_1996}, GGA+U~\cite{dudarev_1998}, with $U$ (Cu-\emph{d}) = 5\,eV, and the strongly constrained and appropriately normed (SCAN) meta-GGA~\cite{sun_2015} using the projector augmented wave approach (PAW) implemented in the VASP code~\cite{kresse_1999}.  The present results were obtained from collinear spin-polarized calculations including relaxation of atomic forces and cell volume and shape with a stopping criterion of 0.03 eV/Å. For the 41 atom cell, we use a 2×2×4 k-mesh for Brillouin sampling. All three functionals yield broadly similar results.

Within the 41 atom unit cell, Cu substitution on the original Pb(4\emph{f}) site leads to 4 different configurations, whereas there are 2 non-equivalent substitutions on Pb(6\emph{h}). Notably, the 4\emph{f} substitutions preserve the P3 symmetry of the 41 atom Pb$_{10}$(PO$_4$)$_6$O cell, but the 6\emph{h} substitutions result in a structure without remaining symmetries. However, even when the initial atomic structure preserves some symmetries, open-shell systems tend to break these symmetries to open an energy gap between occupied and unoccupied spin and crystal-field orbitals, thereby lowering the total energy. This phenomenon can be described as a cooperative Jahn-Teller effect, where the electronic symmetry breaking coincides with or arises from a symmetry breaking in the atomic structure~\cite{gehring_1975}.  Such symmetry breaking is often suppressed in DFT calculations when the charge density is symmetrized to reduce the computational effort. In order to account for a possible cooperative Jahn-Teller effect, we also perform calculations without the symmetry constraint. To lift the initial degeneracies, we apply stochastic perturbations in the initial atomic and/or electronic structures. Specifically, we add random atomic displacements up to 0.03~{\AA}, or preserve a degree of randomness in the initial wave function before starting the self-consistency cycle. Either approach, or a combination of both, is suitable to find symmetry-broken low-energy solutions.

Table~\ref{tab:1} summarizes the relative energy differences between the different configurations for Cu substitution in 41 atom unit cell, as well as energy gains resulting from the symmetry breaking. In all cases, including the symmetry constrained ones, the self-consistency cycle converges to a total magnetic moment of 1\,$\mu_B$/Cu. The four 4\emph{f} site substitutions result in 2 pairs with almost identical total energies within each pair. Therefore, only 2 energies are given for the 4\emph{f} sites in Table~\ref{tab:1}. One 4\emph{f} pair results in the lowest overall energies in the GGA+U and SCAN functionals. Since SCAN likely provides the most accurate total energies, we use this configuration, labeled 4f-1, as the reference energy in~Table~\ref{tab:1}. The energies obtained with symmetry-constrained calculations are higher by $\Delta E \ge 0.4$~eV/fu indicating a very substantial energy gain resulting from the Jahn-Teller distortion. 

In the GGA calculation for the 6\emph{h}-2 substitution, we observe an interesting effect in that the atomic relaxation, starting from the GGA+U relaxed structure, approaches a local minimum, but then picks up larger forces again and settles in a new local minimum. Feeding this structure (labeled 6\emph{h}-2’ in~Table~\ref{tab:1}) back into a GGA+U relaxation leads to a lower energy than the original 6\emph{h}-2. Even though this minimum remains above the energy of 4f-1 in GGA+U and SCAN, this observation is significant as it indicates that Cu substitution in Pb$_{10}$(PO$_4$)O$_6$ could result in numerous local minima with different types and degrees of local distortions. Such complex energy landscapes could result in strong electron-phonon coupling effects with potentially important implications for superconductivity.

\begin{table}
\centering
\caption{Relative energy differences (eV/fu) between different Cu configurations in the 41 atom unit cell of CuPb$_9$(PO$_4$)$_6$O, calculated in the GGA+U, GGA, and SCAN functionals. Given are the energies for the symmetry-broken (sb) gapped state, and for the half-metallic symmetry-constrained (sc) calculation in case of the 4f site substitutions. Generally, the sb state with Cu on the 4f-1 site is the most favored.  \\}

\begin{tabular}{|l|cc|cc|cc|cccc}
\hline
 &GGA+U& &GGA& & SCAN&\\
\hline
Cu site&sb&sc&sb&sc&sb&sc\\
\hline
4f-1&0&0.83&0&0.41&0&0.46\\
4f-2&0.08&1.23&0.08&0.65&0.27&1.04\\
6h-1&0.13&&0.04&&0.24&\\
6h-2&0.11&&&&0.50&\\
6h-2'&0.04&&-0.03&&0.44&\\
\hline
\end{tabular}
\label{tab:1}
\end{table}

\section{Results: Electronic structure properties}

Without Cu substitution, Pb$_{10}$(PO$_4$)$_6$O is an insulator with a gap of several electron volts (see~Fig.~\ref{fig:2}). Determining the precise magnitude of the gap does not lie within the scope of the present work, as it requires beyond-DFT methods and consideration of spin-orbit coupling within the unoccupied Pb-6p manifold. Upon Cu substitution on the Pb(4f) sites with the constraint of preserving the initial symmetry in the charge density, the system becomes a half-metal (the band structure features will be discussed in more detail below). However, allowing for a breaking of the P3 symmetry causes an energy gap to open, consistent with Cu$^{2+}$ in a $d^9$ configuration. A gap also opens in case of the Pb(6\emph{h}) substitution, where the symmetry is already broken in the initial atomic structure. Notably, numerous preprints reported half-metallic band-structures on the basis of DFT calculations and failed to observe the opening of a gap~\cite{griffin_2023, lai_2023, hao_2023, si_2023, cabezasescares_2023, tao_2023, jiang_2023}, presumably due to the above described symmetry constraint.  The gapless band structure was also used for construction of tight-binding type models for this material~\cite{tavakol_2023, oh_2023}.

Table~\ref{tab:2} gives the (generalized) Kohn-Sham energy gaps for the different functionals and atomic configurations described above. While the gaps of the symmetry-constrained calculations (not shown in~Table~\ref{tab:2}) are zero for the DFT functionals considered here, we note that even with the symmetry constraint in atomic structure and charge density, a gap opens up at a higher level of theory, i.e QSGW calculations. These effects will be discussed in more detail in future work. Here, all configurations without explicit symmetrization show sizable band gaps of comparable magnitudes, with the GGA+U and SCAN functionals giving larger gaps than GGA. It is remarkable that even standard DFT at the GGA level predicts a gap opening (Table~\ref{tab:2}), despite the well-known failure to predict any gaps at all in many Mott insulators, e.g., in CuO~\cite{wu_2006}.  It must be expected that beyond-DFT methods, such as many-body perturbation theory~\cite{hedin_1965}, result in substantially larger gaps.

\begin{table}
\centering
\caption{Energy gaps (eV) of the different Cu configurations in 41 atom unit cell of CuPb$_9$(PO$_4$)$_6$O, as obtained from the (generalized) Kohn-Sham energies for the different considered functionals, all calculated for the symmetry broken case. \\}

\begin{tabular}{lrrr}
Cu site&GGA+U&GGA&SCAN\\
\hline
4f-1&1.22&0.56&1.31 \\
4f-2&1.12&0.46&1.20\\
6h-1&1.42&0.75&1.46\\
6h-2&1.17& &1.25\\
6h-2'&1.19&0.64&1.29\\

\hline

\label{tab:2}
\end{tabular}
\end{table}

The gap opening requires the formation of a local magnetic moment at the Cu site. All gapped solutions have indeed a strong local moment, about $0.7$~$\mu_B$ in GGA+U and SCAN, and slightly smaller at about 0.6 ~$\mu_B$ in GGA (magnitude determined from PAW projection). The local moment in the half-metallic symmetry-constrained cases is slightly smaller, but still comparable, e.g., 0.70 vs 0.72~$\mu_B$ for 4f-1 in SCAN. The reason for the metallicity is that the symmetry constraint enforces the degeneracy of the two highest crystal field orbitals in the local Cu minority spin channel, whereas the symmetry breaking lifts the orbital degeneracy. Thus, the present results are consistent with a Cu$^{2+}$~$d^9$ configuration. Such a configuration is typically unstable against on-site electronic correlations that can also contribute to gapping, i.e. we can also consider these to be in the Mott family (or more precisely the charge-transfer insulator family, following the classification of Zaanen, Sawatzky, and Allen~\cite{zaanen_1985}. In the present case (as in most) both the structural (Jahn-Teller like) and correlation (Mott like) seem to be active. 
In order to test whether the atomic and electronic (spin and orbital) ordering, as implied by the periodic boundary conditions of the small 41 atom cell, affects the magnitude of the local moment and the energy gap, we performed supercell calculations. First, constructing a 1×1×2 supercell of Pb$_{10}$(PO$_4$)$_6$O with 82 atoms, we relaxed all 7 non-equivalent configurations with 2 out of the 8 O(4) sites occupied. Using the lowest energy structure, we then generated 6 random Cu pair configurations on the original Pb(4f) sites and calculated for each the ferromagnetic (FM) and anti-ferromagnetic (AF) Cu spin configurations in GGA+U. Taking, again, the lowest energy configuration, we also calculated the respective SCAN energies. Here, the overall energy lies 0.27 eV/fu below that of 4f-1 in the 41 atom cell, indicating that the small cell models discussed above do not provide the absolute global energy minimum, although probably a reasonable approximation thereof. The AF configuration has a slightly lower energy than FM with $E_{AF-FM} = -1.7$~meV/Cu. The small energy difference indicates that the system is likely paramagnetic (spin disordered) down to low (cryogenic) temperatures. The Cu local spin moments are virtually unchanged (0.72~$\mu_B$) in both AF and FM and the band gaps increase slightly (by ~0.2 eV), when compared to the 4f-1 configuration of the 41 atom cell (cf.~Table~\ref{tab:2}). 

As an additional check, a quasirandom spin configuration of the 41 atom 4f-1 configuration was calculated in the 328 atom supercell, and the lattice was allowed to relax. As in the 82-atom AF case, the local Cu moment and the bandgap were nearly constant, and lattice relaxation had only a minor effect. To test disorder effects in the large 328 atom supercell, we calculated a configuration with 1 Cu atom at an alternative Pb(4f) site, giving similar results. Significant relaxation of the structure was needed, mostly visible in modification of the distance of Cu and Pb defect atoms. However, we invariably find that AF configurations maintain with local magnetic moments of similar magnitude at all Cu atoms, with a non-degenerate $d^9$ orbital configuration. Even in the absence of more systematic configuration sampling, these supercell results strongly indicate that the local Cu moments and the gap opening are robust against atomic disorder (Cu distribution on Pb sublattice and O distribution on fractionally occupied O(4) sites) and against Cu spin fluctuations.   

\begin{figure*}
\centering\includegraphics[width=1.0\linewidth]{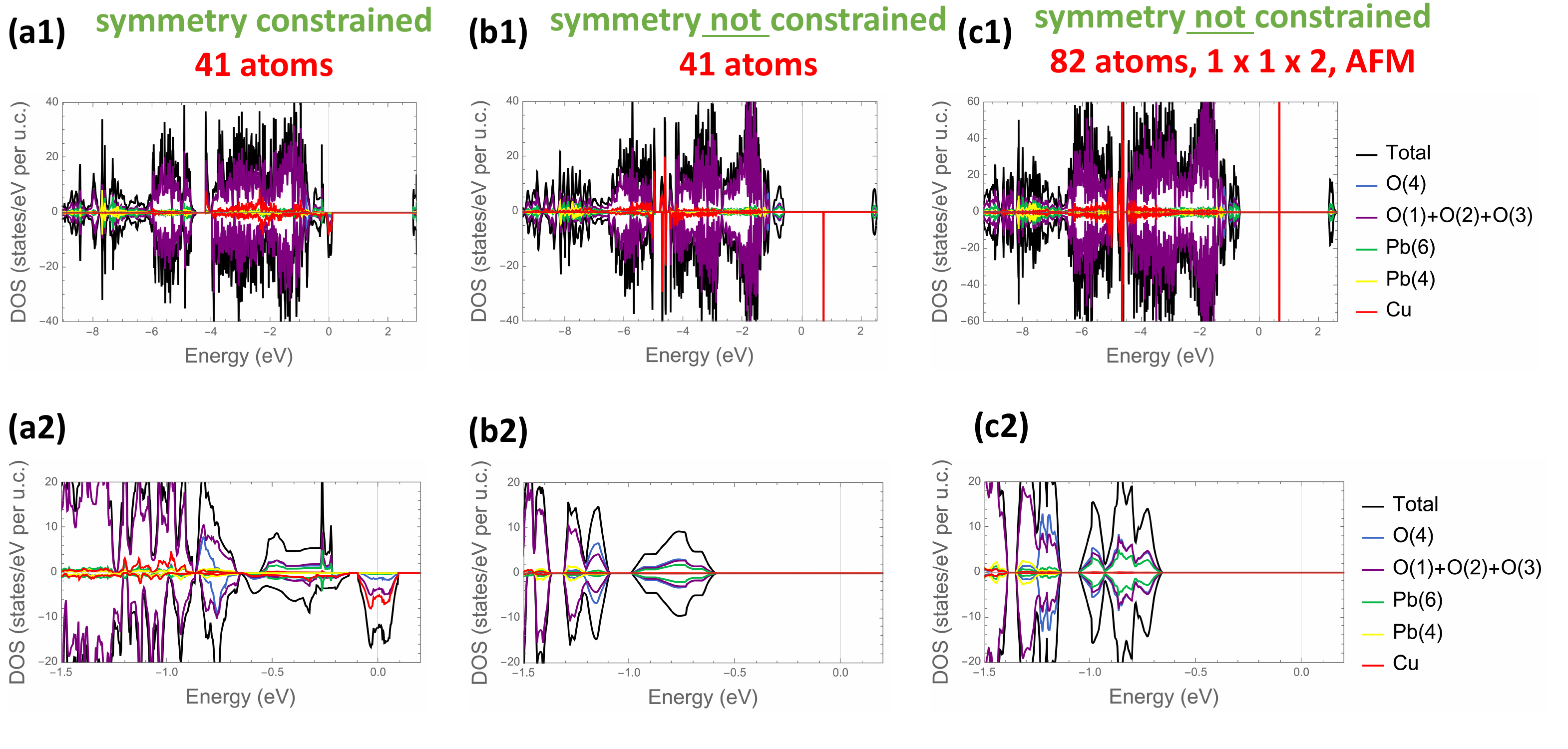}
\caption{Density of states of CuPb$_9$(PO$_4$)$_6$O calculated with the GGA PBE functional and $U=5$~eV at the Cu
site calculated  for different unit cell sizes and using different symmetry options. Negative and positive values denote opposite spin directions. (a1) and (a2) Calculations with the original P3 symmetry constrained. A narrow set of bands of dominant Cu-O origin appears at $E_F$, with no gap indicative of a metallic state. The electron count is near $d^9$, similar to the cuprate superconductors. (b1) and (b2) Results of calculations preformed without symmetry restrictions. Total energy is lowered and a gap will open at the Fermi level, producing an insulator (see~Table~\ref{tab:1}). (c1) and (c2) Calculations with unrestricted symmetry but in a $1\times1\times2$ supercell allowing for AFM configuration for out-of-plane spins. Despite the different magnetic order the DOS close to $E_F$ is qualitatively similar to the ones presented in panels (b1) and (b2). }
\label{fig:4}
\end{figure*}

We have calculated density of states (DOS) of CuPb$_9$(PO$_4$)$_6$O for each supercell size discussed above   using GGA PBE+U functional with $U$=5\,eV on Cu \emph{d} shell. In Fig.~\ref{fig:4} we show only results for selected   41 atom and 82 atom cells as the other ones are expected to be qualitatively similar. We compare results of two different   calculations. Firstly, we imposed P3 symmetry for both, crystal structure and electronic degrees of freedom (panels (a1) and (a2)). The other calculations were performed without symmetry restrictions (panels (b1),(b2),(c1) and (c2)). In both cases, we   considered only collinear magnetism with initial magnetic moment $\sim 1$~$\mu_B$ entirely on Cu atom. The symmetry   restricted case is consistent with a metallic state with large and spin polarized DOS at $E_F$ coming mostly from Cu dopant, with significant admixture from oxygen from PO$_4$ tetrahedra, denoted here as O(1)+O(2)+O(3).

\begin{figure*}
\centering\includegraphics[width=1.0\linewidth]{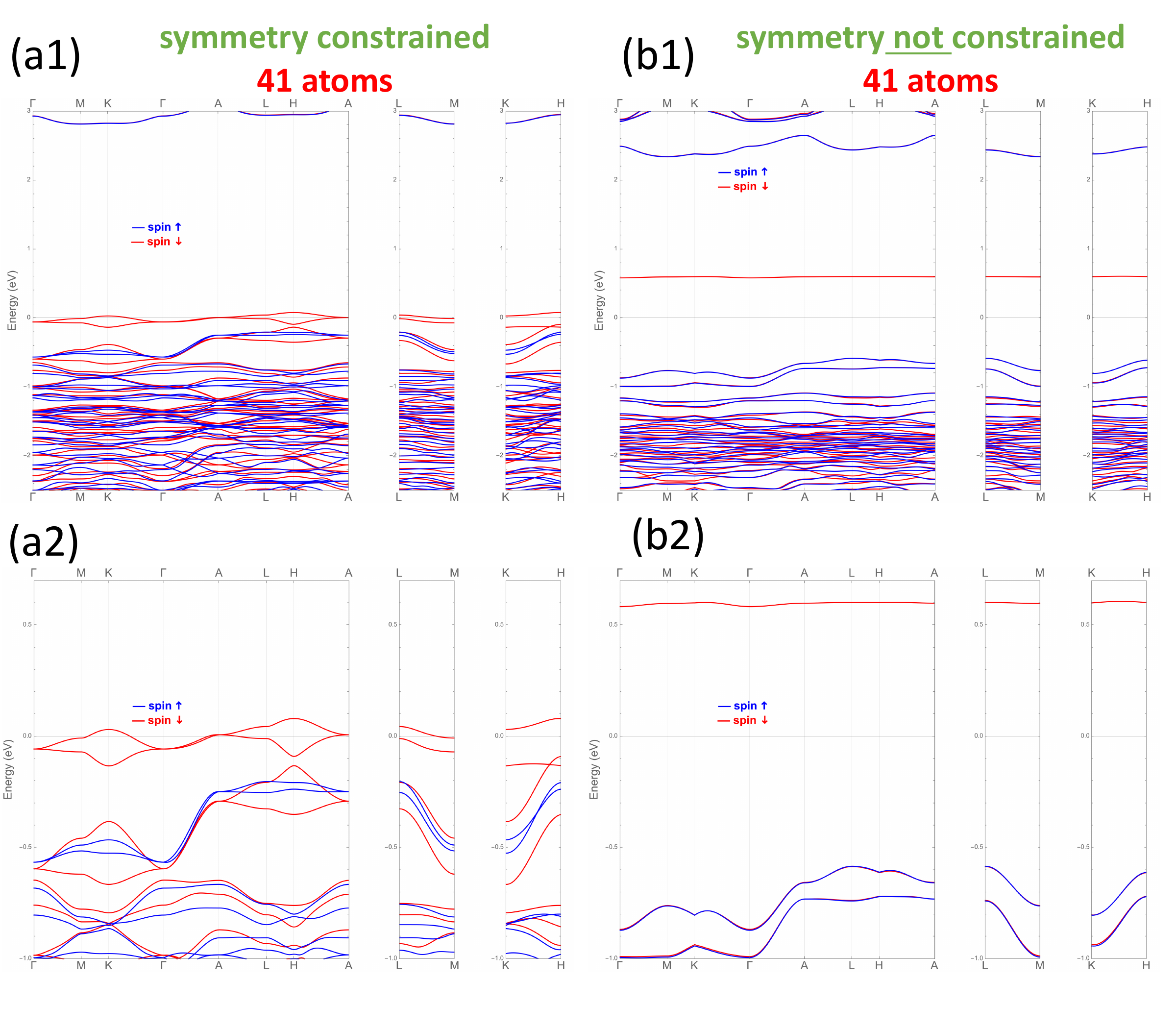}
\caption{The band structure of CuPb$_9$(PO$_4$)$_6$O associated with the DOS of~Fig.~\ref{fig:4}. (a1) and (a2) show the band structure obtained from P3 symmetry-constrained calculations.(b1) and (b2) show the band structure obtained from calculations without symmetry constraints imposed. Only results for 41 atom unit cells are presented. In the case of results presented in panels (b1) and (b2) we use the same path as for (a1) and (a2), ignoring the effects of the weak triclinic distortion.}
\label{fig:5}
\end{figure*}

A different   situation is observed for the symmetry unrestricted calculations~(panels (b1) and (b2)). The energy gap opens around $E_F$ for the symmetry relaxed cases and in this case the top of the valence band can be decomposed in comparable contributions of O(4) and O(1)+O(2)+O(3). The next leading contribution comes from Pb(6) sites. Cu states are now visible as a sharp peak in DOS at $\sim1.2$~eV above the top the valence band. Interestingly, this behavior is preserved for $1\times1\times2$ supercell with AFM spin arrangement (panels (c1) and (c2)), showing the Cu moment is a function only of its local environment. It is noteworthy that the relaxation of the symmetry constraint changes the character of the states around the Fermi level. Whereas in the symmetry-constrained case, $E_\mathrm{F}$ intersects the metallic bands with predominant Cu characters (see Fig. \ref{fig:5}~a), in the symmetry broken case, $E_\mathrm{F}$ lies between the empty Cu band and the occupied "valence band" states, which have Pb(6\emph{h}) and oxygen character from all O sites, including O(4). In all   considered cases we get filling of Cu \emph{d} states equal to 9, similarly to the parent compounds of traditional cuprate superconductors.

Fig.~\ref{fig:5} shows $E$ vs. $k$ dispersion relations for the different calculations. Panels (a1), (a2) correspond
to the P3 symmetry preserving calculation, while panels (b1) and (b2) correspond to the symmetry unconstrained case. We do not
present bands for any other unit cell size than 41 atoms, because it would need band unfolding. In case of P3 preserving calculations one can see a set of narrow ($< 0.2$~eV)
 bands of dominant Cu-O origin at $E_F$, with no gap, indicative of a metallic state. The system is
half-metallic and has a net magnetic moment of 1~$\mu_B$ with ~0.7 of it localized on the Cu site in DFT.  In the
simplest 41 atom cell the Cu are ferromagnetically aligned and the two spin channels are different, giving rise to a
half-metal. The conduction band is found only in the minority channel with a single, weakly dispersive ($\approx 0.2$~eV bandwidth) state mostly
of Cu character forming the conduction band, and the valence band consisting of mostly O and Pb derived states. We noticed that the flattest dispersion close to $E_F$ is found around the $L$ point, in the AL direction. The small electron pocket with $k_F=0.22$~1/{\AA} and the bottom at $\varepsilon=-9.9$~meV is characterized by effective mass $m^{\star}=18 m_e$. The same effective mass can can be found the $\Gamma$ plane, however the flattest part of the band is well below the Fermi level ($\sim 60$~meV). 

The band structure is changed significantly when symmetry constraints are not imposed (panels (b1) and (b2)) This is despite the fact that a triclinic distortion of the unit cell is relatively small, so the deviation from the hexagonal Brillouin zone can be basicaly ignored. The flat Cu band got pushed well above the Fermi level ($\sim1.2$~eV). The width of this band is less than 24~meV. This band belong to "spin-down" spin down manifold. Its spin polarization is in contrast with occupied bands which have their opposite spin counterparts extended roughly over the same binding energy range.

\begin{figure*}
\centering\includegraphics[width=0.9\linewidth]{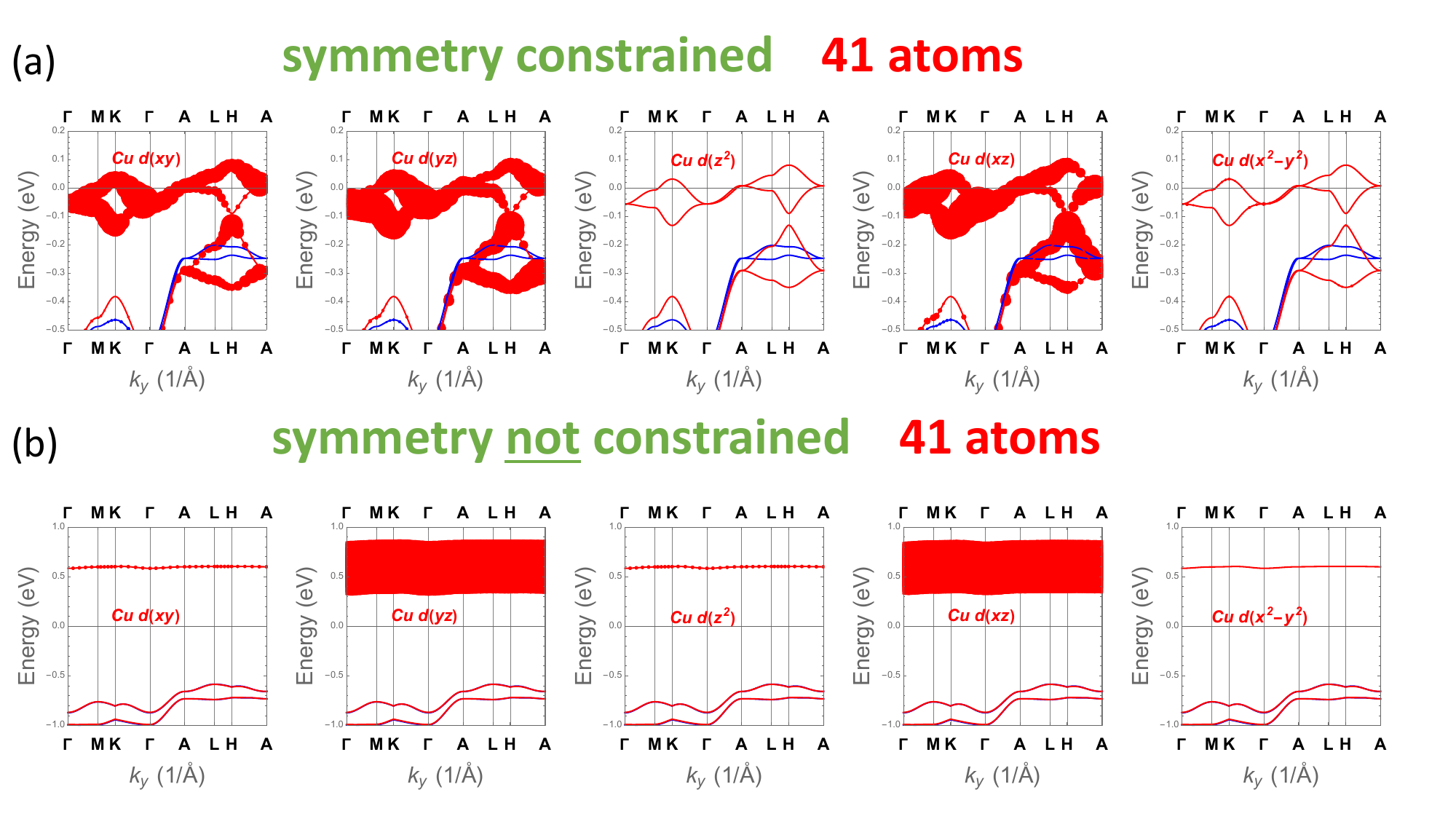}
\caption{Orbital projections of the flat bands at $E_F$ in CuPb$_9$(PO$_4$)$_6$O, with spectral weight indicated by the size of the circles. In contrast to standard cuprate superconductors which have the near $E_F$ weight from $d(x^2-y^2)$ orbitals, the  dominant weight in the wave function comes here from the $d(xy)$, $d(xz)$, and $d(yz)$ orbitals in the symmetry constrained case (a). Increase in directionality is visible in case of calculations without symmetry restrictions (b). For the symmetry-unconstrained case the contribution is almost fully from  $d(xz)$, and $d(yz)$ states.}
\label{fig:6}
\end{figure*}

Fig.~\ref{fig:6} shows projections of the calculated bands onto the various Cu d-orbital symmetries, again for two different types of calculations: preserving P3 symmetry (panel a) and with no symmetry constraints (panel b). In the first case, we find the the
$d(xy)$, $d(xz)$, and $d(yz)$ states all have significant spectral weight at $E_F$, in strong contrast to conventional
cuprate superconductors where the bands at $E_F$ are $d(x^2-y^2)$ derived.  This difference has to do with the local
bonding directionality between the Cu and O atoms, though it is not clear yet what impact, if any, this might have on
superconductivity. The orbital polarization is even stronger in the second case (no symmetry constraints). The contribution of $d(xy)$ is strongly suppressed in the first  conduction band ($\sim$0.5 eV above the nominal $E_F$, which is put in the middle of the gap). This implies even stronger directional hybridization effects.

There are strong similarities with undoped conventional cuprates: the closest Cu-O bond is similar, and states near the Fermi level are predominately of Cu and O character.  Moreover, the system is AF in the ground state with significant local Cu momemts, contributing to the gap opening.  Unlike the cuprates, the valence band in the relaxed structure consists of a pair of mostly O-derived bands with little Cu character.  Also the O-derived valence bands and the Cu-derived conduction band are very flat. The Cu bands are essentially dispersionless, but also O-derived valence bands in the Brillouin zone in the plane normal to $k_z$ are very flat along in-plane directions. That implies a very large part of the Brillouin zone at~$E_F$ has extremely flat bands.

\section{Discussion and implications}
Our finding, through calculation, of flat Cu-O bands at $E_F$ in Cu-doped Pb apatite show that this kind of system can be promising for superconductivity, though a critical aspect is whether or not the system will remain metallic, which is of course a requirement for supporting superconductivity. Our calculations indicate both challenges as well as possible paths for overcoming these challenges. First, we show that the system with exactly one Cu per cell (x=1) will be a metal only if the structure can maintain its relatively high P3 symmetry - and that it is generally unstable to a Jahn-Teller style symmetry lowering that opens a substantial gap at $E_F$, turning the system into a charge-transfer insulator, which likely also has Mott-style correlations. Even though the structural changes are very modest between the symmetry-preserved and symmetry-lowered systems (bond angle changes maximum 1 degree), the electronic structure of the two are quite different, with the near-$E_F$ states of the latter showing significantly less Cu-O hybridization. Therefore, to support a metallic (and possibly superconducting) ground state in this system, at least one of two things must occur: (1) Some internal structural pressure must exist to maintain the higher P3 symmetry of the system, or equivalently, to disallow the energy-lowering distortion. This could in principle "protect" the flat bands at $E_F$, keeping them available for supporting superconductivity. (2) A departure from the "pure" stoichiometry considered here would be necessary, for example by varying the concentration of O(4) atoms away from unity. Since the O(4) atoms are nominally 2- valence, a surplus of such O(4) atoms would act as a hole dopant or a deficit acting as an electron dopant. This could either lead to new impurity-like states within the band gap, or could shift the Fermi energy into the Cu-derived conduction band (electron-doped) or the Pb-O derived valence band (hope-doping). Such a scenario could potentially lead to a metallic flat-band state with interesting properties that should be explored by future experimental and theoretical studies. 
 
Assuming that such flat bands and their associated high density of states can be stabilized by the mechanisms discussed above, such bands could be strongly favorable for superconductivity. This is because bosons that might mediate electronic pairing are in general conducive to strong correlations from flat bands: for example the electron-phonon interaction can be greatly enhanced, and superconductivity mediated by spin fluctuations are also greatly enhanced by the flatness of the band; see for example Ref.~\cite{acharya_2023}. Extremely flat states at the Fermi energy can also significantly enhance the electron-phonon coupling matrix elements, which would also support higher superconducting transition temperatures. In contrast with the hydride superconductors, the phonon modes are less violent for oxygen and copper and may not need high pressures to prevent the system from falling apart. The presence of disorder and flat electronic states together suggest the material may choose an s-wave superconducting phase. Most importantly, the Cooper pairing in that case will probably be better described by a strong coupling picture akin to the BEC limit where $T_c/T_F > 1/10$.

However,  superconductivity is only one of many possible instabilities.  It must compete with, e.g. magnetically ordered phases.  An excellent illustration of this is CaFe$_2$As$_2$ : which is dominant depends on the details of the lattice structure~\cite{acharya_2020}.  Similarly, charge density waves can compete with the electron-phonon interaction, or possibly enhance it.  It requires a more careful analysis of all possible competing instability-channels to explore what the material chooses as its ordered ground state. 
To summarize, we turn to~Fig.~\ref{fig:1}. We have established that the weak hybridization and flat bands can lead to instabilities of different kinds (a) and (b).  The pristine compound is a charge-transfer insulator, and appears to be robust against the disorder we have considered here.  How a metallic phase forms from the near-$E_F$ states needs to be resolved by a combination of theory and experiment.  The question of which boson predominates for correlated phenomena, including superconductivity, may well depend on the character of the metallic state. \\

\hrule

Note added:  We are aware that works describing DFT studies of LK-99 have been posted on aXiv at similar time as our submission~\cite{griffin_2023,lai_2023_1}. We also need to note that need for a doping to achieve conductive state has been suggested very recently on the basis of DMFT calculations~\cite{korotin_2023,si2023_2,yue_2023}. Prediction of half-metallic band structure for LK-99 has also been posted very recently~\cite{hao_2023_1}. A paramagnetic LK-99 that is not superconducting has also been realized in experiment~\cite{Kumar_2023}.Recent experimental study of LK-99 states that room temperature superconductivity is very unlikely in this system~\cite{puphal_2023} and observed transition might be related to the Cu$_2$S impurity phase~\cite{zhu_2023}.

\section{ ACKNOWLEDGMENTS}
We thank Dushyant Narayan for help in setting up some of the calculations. Work at the University of Colorado at Boulder was supported by the U.S. Department of Energy, Office of Science, Office of Basic Energy Sciences, under Grant No. DE-FG02-03ER46066, and the Gordon and Betty Moore Foundation’s EPiQS Initiative through Grant No. GBMF9458. This work was authored in part by the National Renewable Energy Laboratory, operated by Alliance for Sustainable Energy, LLC, for the U.S. Department of Energy (DOE) under Contract No. DE-AC36-08GO28308.  DP, SL, SA and MvS were supported by Office of Science, Basic Energy Sciences, Division of Materials. Calculations were performed using computational resources sponsored by the Department of Energy: the Eagle facility at NREL, sponsored by the Office of Energy Efficiency and also the National Energy Research Scientific Computing Center, under Contract No. DE-AC02-05CH11231 using NERSC award BES-ERCAP0021783. The views expressed in the article do not necessarily represent the views of DOE or the U.S. Government. The United States Government retains and the publisher, by accepting the article for publication, acknowledges that the United States Government retains a non-exclusive, paid-up, irrevocable, worldwide license to publish or reproduce the published form of this manuscript, or allow others to do so, for United States Government purposes.

\bibliographystyle{apsrev4-1}
\bibliography{LK99_refs}









\end{document}